\documentclass[aoas,nameyear,dvips]{arximspdf}

\doi{10.1214/00-AOAS312C}
\referstodoi{10.1214/09-AOAS312}
\volume{3}
\issue{4}
\pubyear{2009}
\firstpage{1279}
\lastpage{1281}

\begin{document}
\begin{frontmatter}

\title{Discussion of: Brownian distance covariance}
\pdftitle{Discussion on Brownian distance covariance by G. J. Szekely and M. L. Rizzo}
\runtitle{Discussion}
\begin{aug}
\author[A]{\fnms{Leslie} \snm{Cope}\ead[label=e1]{lcope@jhsph.edu}\corref{}}
\runauthor{L. Cope}
\affiliation{Johns Hopkins University}
\address[A]{The Sydney Kimmel Comprehensive\\\quad Cancer Center\\ Johns Hopkins University\\ Baltimore, Maryland 21205\\
USA\\
\printead{e1}} 
\end{aug}




\end{frontmatter}

I read \textit{Distance Covariance}, by Drs. Szekely and Rizzo, with
great interest. This is an elegant contribution to statistical theory;
the three-way equivalence between a weighted expectation of the
difference between Brownian covariance and two very different
formulations of ${\mathcal V}^2$ is very attractive, and together with
the examples make a strong case for distance covariance.

But like many statisticians, I spend much of my working life analyzing
genomic data sets and so am interested in how distance covariance and
correlation might be used in high dimensional data with relatively
small sample sizes. In these applications it is often more important to
characterize the relationships between genes than to formally test for
independence. And the Pearson correlation coefficient, complimented by
a well-developed and widely-used theory of linear models and matrix
methods, is highly applicable on such data sets. The restriction to
linear relationships between variables is arguably even an advantage;
while Pearson's correlation may not capture all dependencies, we know a
great deal about the interpretation of results from its application.

It is, of course, not possible to settle the question here, but some
preliminary thoughts follow on the potential utility of distance
covariance, and particularly the scaled distance correlation, in this setting.

Using the author's notation, if $(X,Y)$ is a pair of random variables
(vectors) and $(\mathbf{X}, \mathbf{Y})$ a sample drawn from the joint
distribution, the dependence statistics $A_{kl}$ and $B_{kl}$ are
centered, interpoint distance matrices for $\mathbf{X}$ and $\mathbf{Y}$
respectively, and ${\mathcal V}^2(\mathbf{X},\mathbf{Y})$ is the mean product
moment of the entries in these two matrices.
Thus, the empirical distance covariance is a cross-variable covariance
of within-variable interpoint distances, and the distance correlation
is the same, appropriately scaled. In practice, this is similar to the
\textit{correlation of correlations} used by Lee et al. (\citeyear
{Leeetal2003}) and Parmigiani et al. \citeyear{parmetal2004} to quantify
the reproducibility of results obtained on different microarray
platforms or from independent gene expression studies, but is more
general, since it can be applied even to two, scalar-valued random
variables, and because of its potential to capture nonlinear as well as
linear dependencies.\looseness=1

This representation of the distance correlation offers some intuition
into the characteristics of the statistic. It is reflected in Theorem
4(iii), stating that $R(\mathbf{X},\mathbf{Y})=1$ only if $\mathbf{Y}$ can be obtained
from $\mathbf{X}$ by orthogonal transformation, since these rigid transformations
preserve interpoint distances up to a scaling factor. It explains the
ability, demonstrated in the first few examples, to capture
nonmonotone relationships between two variables; samples with similar
wavelength also have similar transmittance. It also helps to explain
why the method does not offer an advantage over Pearson correlation
when the relationship between the variables is monotone if nonlinear,
as in the example of Gumbel's bivariate exponential random variables.
In the case of monotone dependence, there is much less difference
between correlating interpoint distances and correlating the original variables.

This representation also sheds light on one property of the empirical
estimate of distance covariance that may be very important in the small
sample, high dimensional setting typical of genomic studies. While
shown to be consistent, it is not unbiased. For small and even moderate
sample sizes, there can be a substantial bias, increasing with the
dimensionality of the data. Suppose that $(\mathbf{X}, \mathbf{Y})$ is a small
sample drawn from the joint distribution of $X$ and $Y$. If $i \neq j$,
then the Euclidean distance between $\mathbf{Y}_i$ and $\mathbf{Y}_j$ is a
random value with distribution depending on the variance of $Y$, and if
$i=j$, then the distance is 0. Even after the centering step, the
distribution of values on the diagonal of each distance matrix is very
different from that found off-diagonal, and contributes to inflated
distance covariances and correlations. As the sample size increases,
the influence of the diagonal decreases, and so this source of error
vanishes in the limit.

The same bias affects other potential applications of the method.
Principle components analysis has many applications in genomic data
analysis and one might apply the same decomposition to a matrix of
pairwise distance covariances or correlations. The consistent inflation
of these quantities for every pair of variables puts significant load
on a spurious component, depending on the variance of each variable.

This does not present problems for a permutation test of independence
based on this statistic, where the null distribution exhibits the same
bias, and the lack of power that goes with it is not unexpected when
the sample size is small. It may be that simply excluding the diagonal
elements of the distance matrices from the final covariance calculation
makes for a reasonably unbiased, finite sample estimator, but for the
version presented in this paper, this does complicate interpretation,
and may invalidate parametric, asymptotic tests.

I strongly suspect that the authors are right when they say, ``In
summary, distance correlation is a valuable, practical, and natural
tool in data analysis and inference$\ldots,$'' but believe that potential has
not yet been fully demonstrated, and look forward to further
developments that may do so.

\printaddresses

\end{document}